\documentclass[english]{RBIEarticle} 

\usepackage[utf8]{inputenc} 
\usepackage[T1]{fontenc} 

\usepackage{colortbl}
\definecolor{gray}{gray}{.8}

\usepackage[cmex10]{amsmath}
\usepackage{tikz,pgfplots}
\usepackage{algorithm,algorithmic}
\usepackage{amssymb}
\usepackage{scalefnt}
\usepackage{adjustbox}

\title{New Metrics for Learning Evaluation in Digital Education Platforms}

\author{%
	\parbox{5.5cm}{%
		Gabriel Leitão\\
		Federal Institute of Amazonas\\
		gabriel.leitao@ifam.edu.br\\
	}
	\parbox{5.5cm}{%
		Juan Colonna\\
		Federal University of Amazonas\\
		juancolonna@icomp.ufam.edu.br\\
	}
	\parbox{5.5cm}{%
		Edwin Monteiro\\
		Federal University of Amazonas\\
		edwin@icomp.ufam.edu.br \\
	}
	\parbox{5.5cm}{%
		Elaine Oliveira\\
		Federal University of Amazonas\\
		elaine@icomp.ufam.edu.br\\
	}
 	\parbox{5.5cm}{%
		Raimundo Barreto\\
		Federal University of Amazonas\\
		rbarreto2@icomp.ufam.edu.br\\
	}
 }

\Submission{dd/Mmm/yyyy}
\Camera_ready{dd/Mmm/yyyy}
\First_round_notif{dd/Mmm/yyyy}
\Edition_review{dd/Mmm/yyyy}
\New_version{dd/Mmm/yyyy}
\Available_online{dd/Mmm/yyyy}
\Second_round_notif{dd/Mmm/yyyy}
\Published{dd/Mmm/yyyy}

\heading{Leitão, G., Colonna, J. et al.
}{JOURNAL V.vv, N.n – yyyy}

\citeas{Leitão, G., \ldots \& Colonna, J.  (Year). Article title (título em português). Journal), vol(num), pp-pp. DOI: 10.XXXX/XXXX.yyyy.vol.num.pp}

\usepgfplotslibrary{dateplot}
\pgfplotsset{compat=1.17}
\begin{document}
\maketitle

\begin{abstract}
Technology applied in education can provide great benefits and overcome challenges by facilitating access to learning objects anywhere and anytime. However, technology alone is not enough, since it requires suitable planning and learning method ologies. Using technology can be problematic, especially in determining whether learning has occurred or not. Futhermore,if learning has not occured, technology can make it difficult to determine how to mitigate this lack of learning. This paper presents a set of new metrics for measuring 
student's acquired understanding of a content in technology-based education platforms. Some metrics were taken from the literature ``as is", some were modified slighty, while others were added. The hypothesis is that we should not only focus on traditional scoring, because it only counts the number of hits/errors and does not consider any other aspect of learning. We applied all metrics to an assessment conducted in a high school class in which we show specific cases, along with metrics, where very useful information can be obtained from by combining several metrics. We conclude that the proposed metrics are promising for measuring student 
's acquired understanding of a content, as well as for teachers to measure student's weaknesses.

\end{abstract}


\pagebreak


\section{Introduction}\label{sec:introduction}
The school is one of the most important social and cultural institutions, as it provides knowledge integration and socialization. Therefore, schools need not only to insert technological resources into the teaching-learning processes, but also propose new methodologies and pedagogical practices that, integrated with new technologies, help students develop the skills and abilities needed nowadays.
In this case, such challenge is much more complex than simply inserting new technologies.
Using computational tools in the classroom motivates a real review of teaching methods.
Such education redesign makes the learning process more attractive, especially by providing interaction with educational content through games, physical and/or virtual labs or challenges that stimulate problem solving.

Generally speaking, a virtual learning environment (VLE) is any online environment that aims at teaching and learning~\cite{Roman2019}.
VLE is widely used in the context of distance education, because it allows the complete management of a classroom, as well as the use of various media, languages and resources that present information necessary for student learning.
In addition, it has tools that allow for interactions between students and teachers, information exchange, submission of assignments and availability of study content.
Through the VLE, it is possible to track the learning process of students and generate reports on their performance and progress.

The adoption of technology in education gave rise to the term \textit{technology-enhanced learning} (TEL), which describes the application of technology to teaching and learning\cite{Dunn2019}. TEL can engage students with learning and improve knowledge acquisition due to the flexibility of learning it provides. In other words, TEL is any technology that enhances the learning experience.

Technologies make it possible to gather student interaction data throughout the learning process, thereby generating graphs and analysis that support the teacher in both assessment and decision making that involve, for example, the addition of pedagogical activities to reinforce the learning of a subject.
Thus, data obtained from classes and assessments could better infer the learning profile and difficulties of a particular class or specific student. This way, the system can recommend several activities and strategies that are most appropriate and tailored to the real needs of the class or student.

The \textbf{context} of this work is virtual learning environments that use technology-enhanced learning.
We propose that teachers adopt a learning environment to prepare their lessons using digital learning objects such as slides, videos, photos, texts, games, etc., and make assessments within this technology-based educational environment.
Specifically regarding assessment, the platform should include a way to collect additional data at run time, such as the time each student spent on each question and how many times each student changed an answer to the same question.

To generate metrics, the educational platform should also allow the teacher to include other information for each question, such as the topic/subject, the expected time students need to answer the question, the level of difficulty, and the weight of each answer option.
This paper only focuses on assessments based on multiple choice questionnaires using closed-ended questions. Students used devices such as mobile phones, tablets or laptops to access the classes and assessments contained in the virtual learning environment.

Assessing a student is not an easy task~\cite{James2004}. There are several factors that can influence a student's grade, which often go beyond the degree of knowledge about a topic. Sometimes other factors such as nervousness, somnolence, and even lack of basic knowledge about that topic/subject influence student performance. In addition, it is important to act in a timely manner in case of poor student performance.

Despite the several assessment types, we consider \textbf{formative and summative assessment} \cite{Dixson2016}, where the main aim is to educate and improve student performance, not merely to audit it~\cite{Wiggins1998}.
Formative assessment is defined as an activity undertaken by teachers that provides information to be used as feedback to modify teaching and learning activities~\cite{Black2010}.
Summative assessment, on the other hand, is aimed at assessing the extent to which the most important outcomes have been reached at the end of assessment. Unlike formative assessments, which are generally used for providing feedback to students and teachers, summative assessments are generally high-stakes assessments and used to get a final assessment of how much learning has taken place, that is, how much a student knows~\cite{Dixson2016}.
This kind of assessment is typically less frequent, and occurs at the end of instructional segments.
Although teachers can adopt several formative and summative assessment methodologies, all of them are hard to implement, not automatic, and prone to errors. In this paper, we propose a novel method for teachers to measure students performance in such a way that, using the proposed metrics, teachers can adjust instruction to maximize student learning and develop interventions to improve student learning~\cite{Shepard2006}.

Therefore, the following \textbf{problems} were addressed by this paper: \textit{Is it possible to measure student's 
acquired understanding of a content satisfactorily and fairly? Using such metrics, is it possible to take action to mitigate student's poor understanding of a content?}

Current assessments, which we call ``traditional assessment," are binary in the sense that they only assess whether the question is right or wrong. In open-ended questions, besides right/wrong, ``half-right" is also allowed. However, it seems that this is not the most effective and fair way to evaluate a student, simply because this type of evaluation generally neglects several factors, such as: if the student marked the wrong answer but is close to the correct one; the student's degree of doubt; the time taken to solve the question; the student's level of confidence in answering a question; and the estimated level of understanding of a topic.

In this case, when such items are neglected, a good opportunity is missed to alert (or recommend) students about several topics/subjects in which they could improve their performance. Some of these aspects are very easy to implement. For example, simply indicating which topics a student should focus could greatly help them.

This paper presents the following \textbf{contributions}:
\begin{itemize}
    \item a review of the main learning metrics available in the literature;
    \item new learning metrics and implementation of some literature metrics;
    \item experiments with new metrics using a multiple choice assessment from a high school class;
    \item two metrics that better analyze student behavior and identify students who need to receive more attention from the teacher; and  \item  emphasize priority topics for students to study in order to improve the learning quality.
\end{itemize}

\section{Review of metrics in the context of digital education}
\label{sec:related_works}
\label{sec:old_TS}

Traditional learning environments have been modified due to the increased technology applied in education. Thus, the addition of digital devices makes it possible to create new ways for educational assessment.
Technology in education requires new ways to measure knowledge acquisition, individual behavior and the quality of collaboration during activities in virtual learning environments.

Although technology has been used in education, most works still adopt the traditional score as a parameter to measure student learning. However, some works use emoticons~\cite{sales2007uso}, or fuzzy logic~\cite{da2013estudo}, but the assessment is still based on hits or mistakes.


The most common metric used to calculate student performance is \textbf{Traditional Score (\textit{TS})}. 
Traditional score is typically an accuracy measure between the quantity of correct answers ($c$) and the total number of questions ($n$). From this metric, we can define the error rate as the complement of the traditional score by $e=1-TS$. The traditional score and error are normalized in the range $[0, 10]$, according to the Equation~\ref{eq:TS}.

\begin{equation}\label{eq:TS}
TS=10\cdot \frac{c}{n}
\end{equation}

However, we argue that traditional score is not enough to correctly assess student performance. 
Next, we present some metrics found in the literature to evaluate the degree of student knowledge and produce a score from an assessment.

A set of metrics that considers student activity history was described by Biswas and Ghosh \cite{biswas2007novel}, which are: (a) Level of Understanding; (b) Student Learning Rate, described below; and (c) Difficulty level of a subject, topic or concept.

\subsection{Level of Understanding - \texorpdfstring{$L_u$}{Lu}}
\label{sec:level-understanding}

This metric quantifies the relationship between difficulty indices, response time, and deviation. The \textbf{difficulty indices} (topic, concept and question) are described in Table~\ref{tab:Lu}. 
The \textbf{response time} is applied to catch the student's blind guesses and it has to be compared with the expected response time provided by the question author. There are two classes of Response Time, the Blind Guess with a value of 5 and the Normal Answer (or Educated Guess), with a value of 1.
The \textbf{deviation parameter} is given according to the answer classification shown in Table~\ref{tab:Weight_Biswas}, where 0 means ``no idea" and 5 means ``perfect match" or correct answer.

\begin{equation}\label{eq:lu}
L_u = \frac{TDI \cdot CDI \cdot QDI \cdot Deviation}{Response \ Time}
\end{equation}

\begin{table}[htbp]
\centering
\caption{Difficulty Index}
\begin{center}
\begin{tabular}{|l|c|c|c|}
\hline
\textbf{Difficulty Index} & \textbf{Easy} & \textbf{Normal} & \textbf{Hard} \\\hline
Topic Difficulty Index (TDI) & 1 & 3 & 5 \\\hline
Concept Difficulty Index (CDI) & 1 & 3 & 5 \\\hline
Question Difficulty Index (QDI) & 1 & 3 & 5 \\\hline
\end{tabular}
\end{center}
\label{tab:Lu}
\end{table}

Biswas and Ghosh~\cite{biswas2007novel} emphasize that the values of Table~\ref{tab:Lu}, Table~\ref{tab:Weight_Biswas} and Response Time were not derived mathematically, and are only applied to differentiate the distinct student classes. In this case, it is possible to choose any other values.

\begin{table}[htbp]
\caption{Value of deviation parameter}
\centering
\begin{tabular}{|c|c|c|c|c|c|}
\hline
\textbf{Deviation} & \textbf{No Idea} & \textbf{\begin{tabular}[c]{@{}c@{}}Below \\ Average\end{tabular}} & \textbf{Average} & \textbf{\begin{tabular}[c]{@{}c@{}}Near \\ Match\end{tabular}} & \textbf{Match} \\ \hline
\textbf{Value} & 0 & 2 & 3 & 4 & 5\\ \hline
\end{tabular}
\label{tab:Weight_Biswas}
\end{table}

\subsection{Student Learning Rate - \texorpdfstring{$SLR$}{SLR}}

To establish the \textbf{Student Learning Rate} (\textit{SLR}) metric, Biswas and Ghosh~\cite{biswas2007novel} introduced the definition of Student Score, defined by $SS (s,i,o)$, that presents the score of a student $s$ on the $i$-th evaluation in relation to an element $o$, where these elements can be subject, topic or concept.
Thus, student learning rate is the average increase in student score with respect to a set of evaluations. This permits the observation of continuous evolution and is expressed by the Equation~\ref{eq:student_rate}.

\begin{equation}\label{eq:student_rate}
\begin{split}
SLR(s,o) =  \frac{\sum \{(SS(i+1)-SS(i)) \cdot |SS(i+1)-SS(i)| \} }{N-1}
\end{split}
\end{equation}

\noindent where $N$ is the number of questions about the ontology element $o$, and $i$ is a variable expressing a specific assessment.


\subsection{Difficulty Level - \texorpdfstring{$DL$}{DL}}

Biswas and Ghosh~\cite{biswas2007novel} also define that the \textbf{Difficulty Level} (\textit{DL}) of a learning element $o$ (which can be subject, topic or concept) can be quantified for all students $s$ by the Equation~\ref{eq:DL}. Futhermore, \textit{DL} is the \textbf{average} of the Student Learning Rate of all students in all learning elements $o$.

\begin{equation} \label{eq:DL}
DL(o) = \overline{SLR(\forall  s, o)}
\end{equation}



\section{Proposed Metrics}
\label{sec:approach}

As presented in Section~\ref{sec:old_TS}, the existing metrics are only based on hits or proximity to the correct answer, such as the \textit{Traditional Score (TS)} or \textit{Level of Understanding ($L_u$)}, which does not provide any information about student behavior in the assessment.
Therefore, this paper proposes new metrics to overcome such limitation, which complement the evaluation of student learning in a digital education context.

The metrics can be divided in two parts: (i) five isolated metrics; and (ii) three metrics based on the combination of other metrics.

\subsection{Isolated Metrics}

The isolated metrics are: Weighted Score, Question Doubt, Assurance Degree, Student Response Time, and Level of Disorder.

\subsubsection{Weighted Score}
\label{sub:WS}

\textbf{Weighted Score (\textit{WS})} could be a solution to assess students, since we consider that traditional score is not enough to correctly assess student performance, as it only measures mistakes and hits and does not capture when a student almost gets an answer correct or makes a drastic mistake.

This metric is a learning score in the range of $[0,10]$, which is based on the weight ($w_i$) of each selected answer and the maximum weighted punctuation ($mwp$).
Equation~\ref{eq:WS} shows the $WS$ formula, where the weight value of the answers of each question ($w_i$) is an integer ranging from 0 to 4, where 0 is a completely wrong answer, 4 is a completely correct answer, and the others are intermediate answers (see Table~\ref{tab:Weights}).
Obviously, if $w_i$ is equal to 1, 2, or 3, the answer is wrong. However, $w_i = 1$ means \textit{almost} wrong, and $w_i = 3$ means  \textit{almost} correct. This metric is a fairer way of measuring learning performance, since it considers if a student is close to the correct answer. This metric can also be used to suggest what topics to focus on.
The maximum weighted punctuation ($mwp$) is simply the total amount of questions in the questionnaire ($|q|$) multiplied by 4.

\begin{table}[htbp]
\caption{Weight of the answers - w}
\centering
\begin{tabular}{|c|c|c|c|c|c|}
\hline
\textbf{Weight} & \textbf{No Idea} & \textbf{Below Avg} & \textbf{Avg} & \textbf{Near Match} & \textbf{Match} \\ \hline
\textbf{Value} & 0 & 1 & 2 & 3 & 4 \\ \hline
\end{tabular}
\label{tab:Weights}
\end{table}

\begin{equation}
\label{eq:WS}
WS = 10 \cdot \frac{\sum_{i=1}^n w_i}{mwp}
\end{equation}

where 
\begin{description}
\item $w_i$ is from Table \ref{tab:Weights};
\item $mwp = |q| \cdot 4$; and
\item $|q|$ is the total number of questions in the questionnaire.
\end{description}

\vspace{1cm}

Table~\ref{table:studentGrades} is used as an example to explain some metrics proposed in this paper.
 It shows the grades of a student according to weights of the answer options selected in two subjects.

\begin{table}[ht]
\caption{Student Grades}
\centering
\begin{tabular}{|c|c|c|c|c|c|c|c|}
\hline
\multicolumn{8}{|c|}{\textbf{SUBJECT ``A''}} \\ \hline
\textbf{$|q|$ = 6; mwp=24} & \textbf{Q1} & \textbf{Q2} & \textbf{Q3} & \textbf{Q4} & \textbf{Q5} & \textbf{Q6} & \textbf{Sum} \\ \hline
\textbf{w$_j$} & \textbf{4} & \textbf{3} & \textbf{4} & \textbf{2} & \textbf{4} & \textbf{4} & \textbf{21}  \\ \hline
\multicolumn{8}{|c|}{\textbf{SUBJECT ``B''}} \\ \hline
\textbf{$|q|$ = 6; mwp=24} & \textbf{Q1} & \textbf{Q2} & \textbf{Q3} & \textbf{Q4} & \textbf{Q5} & \textbf{Q6} & \textbf{Sum} \\ \hline
\textbf{w$_j$} & \textbf{3} & \textbf{3} & \textbf{3} & \textbf{3} & \textbf{3} & \textbf{4} & \textbf{19}  \\ \hline
\end{tabular}
\label{table:studentGrades}
\end{table}

 In case of Subject ``A", the student missed questions Q2 and Q4.
 Thus, using Traditional Score, the student's grade is 
 $$TS =10\cdot \frac{4}{6} = 6.67$$
 
 However, using the Weighted Score (\textit{WS}), the student does not completely miss all of the questions since the weights ($w_i$) of each question are greater than 0. Therefore, applying Equation~\ref{eq:WS} results in: 
 $$ WS = 10 \cdot \frac{4+3+4+2+4+4}{24} = 8.75 $$
 
Thus, the Traditional Score (\textit{TS}) is 6.67, while the Weighted Score is 8.75. Therefore, WS metric may be a better option for assessing student performance when compared to the Traditional Score (\textit{TS}), since the answers to Q2 and Q4 are not completely wrong.

In case of Subject "B", the student missed questions Q1, Q2, Q3, Q4, and Q5.
Thus, using Traditional Score, the student's grade is 
$$TS =10\cdot \frac{1}{6} = 1.67$$

However, using Weighted Score, the student has:
$$ WS = 10 \cdot \frac{3+3+3+3+3+4}{24} = 7.92 $$

In this case, the discrepancy between \textit{WS} and \textit{TS} is more evident.

\subsubsection{Question Doubt}

\textbf{Question Doubt (\textit{QD})} refers to the number of times that a student returns to the same question and changes it. 
The Question Doubt for question $i$ ($QD_i$) is calculated by Equation~\ref{eq:QD}, where $m_j$  represents these changes on a given question $j$.

\begin{equation}
\label{eq:QD}
QD_i = m_i - 1
\end{equation}

For instance, let's consider a
Current file
Overview
 multiple-choice question. If a student chooses an answer to question $i$ and after a while decides to change it, then $m_i=2$ and $QD_i=1$. However, if the student chooses an answer to question $i$ and does not make any changes, then $m_i=1$ $QD_i=0$. Besides that, if no answer is selected, hence $m_i=0$ and $QD_i=-1$. This allows us to separate the answered questions from the unanswered ones.

\begin{table}[htbp]
\caption{Markings}
\begin{center}
\begin{tabular}{|c|c|c|c|c|c|c|}
\hline
\textbf{Students} & \textbf{Q1} & \textbf{Q2} & \textbf{Q3} & \textbf{Q4} & \textbf{Q5} & \textbf{Q6} \\ \hline
\textbf{Student 1} & 0 & 1 & 0 & 1 & 1 & 1 \\ \hline
\textbf{Student 2} & 1 & 1 & 1 & 1 & 1 & 1 \\ \hline
\textbf{Student 3} & 2 & 4 & 5 & 6 & 8 & 1 \\ \hline
\textbf{Student 4} & 9 & 8 & 6 & 6 & 7 & 9 \\ \hline
\end{tabular}
\end{center}
\label{tab:Markings}
\end{table}

Tables~\ref{tab:Markings} and~\ref{tab:QuestionDoubts} show a sample of ``Markings'' and ``Question Doubts'' for a six-question questionnaire, where the application of Equation~\ref{eq:QD} can be verified in the data from 4 students. In those tables, we can see that Student 1 did not mark questions Q1 and Q3; Student 2 only marked all questions once; and Student 4 marked questions Q1 and Q6 9 different times.

\begin{table}[htbp]
\caption{Question Doubts}
\begin{center}
\begin{tabular}{|c|c|c|c|c|c|c|}
\hline
\textbf{Students} & \textbf{Q1} & \textbf{Q2} & \textbf{Q3} & \textbf{Q4} & \textbf{Q5} & \textbf{Q6} \\ \hline
\textbf{Student 1} & -1 & 0 & -1 & 0 & 0 & 0 \\ \hline
\textbf{Student 2} & 0 & 0 & 0 & 0 & 0 & 0 \\ \hline
\textbf{Student 3} & 1 & 3 & 4 & 5 & 7 & 0 \\ \hline
\textbf{Student 4} & 8 & 7 & 5 & 5 & 6 & 8 \\ \hline
\end{tabular}
\end{center}
\label{tab:QuestionDoubts}
\end{table}

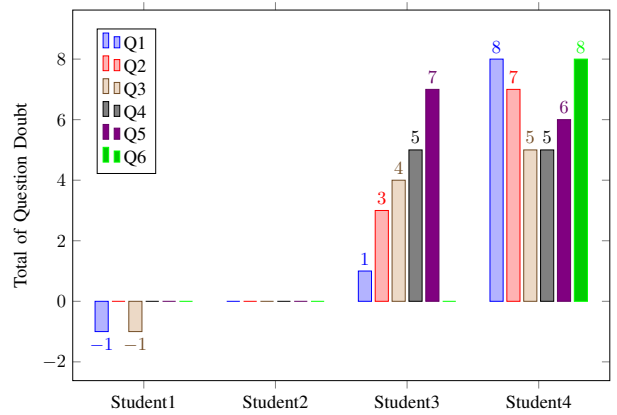
\begin{figure}[ht]
		\centering
		\begin{tikzpicture}[scale=0.70]
		\pgfplotsset{width=10cm,compat=newest}
		\begin{axis}[
		ybar,
		bar width=0.25cm,
		x=2.5cm,
		enlargelimits=0.18,
		legend style={at={(0.1,0.95)},
			anchor=north,legend columns=1},
		ylabel={Total of Question Doubt
		},
		symbolic x coords={Student1, Student2, Student3, Student4},
		xtick=data,
		visualization depends on={y\as\YY},
		nodes near coords
		={\pgfmathtruncatemacro{\YY}{ifthenelse(\YY==0,0,1)}\ifnum\YY=0\else\pgfmathprintnumber\pgfplotspointmeta\fi},
		nodes near coords align={vertical},
		]
		
		\addplot+[] coordinates {
			(Student1,-1) 
			(Student2,0) 
			(Student3,1) 
			(Student4,8)};
		
		\addplot+[] coordinates {
			(Student1,0) 
			(Student2,0) 
			(Student3,3) 
			(Student4,7)};
		
		\addplot+[] coordinates {
		
			(Student1,-1)
			(Student2,0) 
			(Student3,4)
			(Student4,5)};
		
		\addplot+[] coordinates {
			(Student1,0)
			(Student2,0) 
			(Student3,5)
			(Student4,5)};
			
		\addplot+[] coordinates {
			(Student1,0)
			(Student2,0) 
			(Student3,7)
			(Student4,6)};
			
		\addplot+[] coordinates {
			(Student1,0)
			(Student2,0) 
			(Student3,0)
			(Student4,8)};
		\legend{Q1, Q2, Q3, Q4, Q5, Q6}
		\end{axis}
		\end{tikzpicture}
		\caption{Question Doubt}
		\label{fig:QD}
	\end{figure}

In addition, to facilitate data visualization for the teacher, the metric Question Doubt can be depicted as a bar graph (Figure~\ref{fig:QD}).

\subsubsection{Assurance Degree}\label{sec:assurance-degree}

\textbf{Assurance Degree (\textit{AD})} measures the self-confidence of a student when answering a set of questions (from a questionnaire or parts of it) based on the relationship between the total number of correct answers ($c$ - see Section~\ref{sec:old_TS}) and $T$ is the sum of all the answers of each question ($m_i$) of a questionnaire (or parts of it) containing $n$ questions.
Equation~\ref{eq:AD} depicts the $AD$ formula.

\begin{equation}
\label{eq:AD_T}
T = \sum_{i=1}^{n} m_i
\end{equation}

\begin{equation}
\label{eq:AD}
AD = \frac{c}{T}
\end{equation}

For example, given a multiple-choice questionnaire, with three possible answers, if a student chooses the first option and changes it to the second option, and later changes it again to the third option before submitting the final decision, we can infer that this student was not sure about their final choice. This could be because the student is confused about the concepts or is able to self-correct. However, b combining the $AD$ with the $TS$ (Traditional Score - see Section~\ref{sec:old_TS}) or $QuCL$ (Questionnaire Comprehension Level - see Section~\ref{sec:QuCL}), we can observe different student behaviors and, possibly detect which students have more difficulties.

Therefore, Table~\ref{tab:AD} shows some extreme cases that can be analyzed to understand the $AD$ potential. In the first case, suppose that a student never changes their original answers ($c=T$), therefore, their assurance degree is \textit{maximum} ($AD=1= 100\%$). The second case occurs when a student never changes their answers, but all of them are wrong, indicating that they interpreted the concepts incorrectly ($c\approx 0$ or $c\ll T$), in which case $AD$ is the minimum ($AD\approx0$ or $AD\approx0\%$). Finally, the third case characterizes a less problematic behavior in which the student changes their answers many times, but the final score is the maximum  ($c \approx |q|$) indicating an ability to self-correct ($0 < AD < 1$). These cases illustrate the main differences between the simple analysis of the Traditional Score and the new Assurance Degree. Practical examples are shown in Section~\ref{sec:experiments}.

\begin{table}[htbp]
\centering
\caption{Assurance Degree - Extreme Cases}
\begin{tabular}{|c|c|c|c|c|c|c|}
\hline
\multicolumn{7}{|c|}{\textbf{1st Case - AD Maximum}} \\ \hline
\textbf{Question} & \textbf{Q1} & \textbf{Q2} & \textbf{Q3} & \textbf{Q4} & \textbf{Q5} & \textbf{Q6} \\ \hline
\textbf{c} & 1 & 1 & 1 & 1 & 1 & 1 \\ \hline
\textbf{m} & 1 & 1 & 1 & 1 & 1 & 1 \\ \hline
\textbf{AD} & \multicolumn{6}{c|}{100\%} \\ \hline
\multicolumn{7}{|c|}{\textbf{2nd Case - AD Minimum}} \\ \hline
\textbf{c} & 0 & 0 & 0 & 0 & 0 & 0 \\ \hline
\textbf{m} & 1 & 1 & 1 & 1 & 1 & 1 \\ \hline
\textbf{AD} & \multicolumn{6}{c|}{0\%} \\ \hline
\multicolumn{7}{|c|}{\textbf{3rd Case - Self-correction}} \\ \hline
\textbf{c} & 1 & 1 & 1 & 1 & 1 & 1 \\ \hline
\textbf{m} & 4 & 2 & 1 & 3 & 5 & 2 \\ \hline
\textbf{AD} & \multicolumn{6}{c|}{35.29\%} \\ \hline
\end{tabular}
\label{tab:AD}
\end{table}

\subsubsection{Student Response Time}\label{sec:SRT}

The \textbf{Student Response Time (\textit{SRT})} measures the time spent to answer each question or a set of questions. The time spent to resolve specific questions could indicate that the learner did not understand the evaluated topic, the question has a very high difficulty level or it must be rebuilt.

Measuring the response time of a question can be crossed with the Assurance Degree (\textit{AD}) (Section~\ref{sec:assurance-degree}), which may help to visualize the need for further reinforcement of previously studied content or any recommendation to modify the question/questionnaire. For example, a low assurance degree for a given question (labeled as `Easy' by the teacher) associated with a high response time to that question may indicate that the specification of the problem would need to be revised or students did not understand the topic well.

\subsubsection{Level of Disorder}
\label{sec:level-disorder}

The \textbf{Level of Disorder (\textit{D})} is a metric that uses the concept of entropy from the information theory field, which helps quantify the degree of uncertainty of a random variable or the outcome of a random process \cite{Mackay2003book}.
Our main assumption here is that the teacher orders activities in an increasing degree of difficulty, but criterion is subjective and students may experience different degrees of difficulty compared to the teacher.
In other words, the $D$ metric represents the disorder level of the events recorded in a {\it log} file. 

The entropy was originally defined by Shannon~\cite{Shannon1948} based on the probability mass function as:

\begin{equation}\label{eq:H}
H = -\sum_{i=1}^n p_i \log p_i
\end{equation}

\noindent where $p_i$ is the probability of occurrence of the {\it i-th} symbol in a set $N=\{1,2, \dots, n\}$. One of the major problems with this measurement is how to obtain $p_i$ from an unknown probability distribution function (PDF).
We only consider two symbols ($p_1$ and $p_2$) since we only consider two possibilities, namely, \textbf{in order} or \textbf{out of order}.
Thus, we can infer $p_1$ and $p_2$ from the sequence of events as a discrete temporal series, i.e., $p_1$ is the frequency of events that happened in order, while $p_2$ is the frequency of events that were not answered in the expected order.

The pseudo-code of the Algorithm~\ref{algo:PE}.
Consider the following sequence of answers $S = \{ 1a, 2a, 3c, 5b, 1b, 4c, 1a\}$, where the steps carried out to compute the Level of Disorder are:
\begin{enumerate}
   \item to traverse $S$ comparing the number of the correct answer ($s_i$) with the number of the following answer ($s_{i+1}$). Every time that the condition $s_i \leq s_{i+1}$ is satisfied, the counter $p_1$ must be incremented, otherwise the counter $p_2$ must be incremented; 
   \item to normalize $p_1$ and $p_2$ to obtain the probabilities; and
   \item to apply the Equation~\ref{eq:H}.
\end{enumerate}

\begin{algorithm}[H]
	\begin{algorithmic}[1]
		\STATE{$p_1=0;$ $p_2=0;$}
		\FOR{$i=1$ to $n-1$}
			\IF{$s_i \leq s_{i+1}$}
				\STATE{$p_1 = p_1 + 1;$}
			\ELSE
				\STATE{$p_2 = p_2 + 1;$}
			\ENDIF
		\ENDFOR
		\STATE{$D = -\frac{p_1}{p_1+p_2} \ \log \big(\frac{p_1}{p_1+p_2} \big)-\frac{p_2}{p_1+p_2} \ \log \big(\frac{p_2}{p_1+p_2} \big)$}
	\end{algorithmic}
\caption{Level of Disorder}\label{algo:PE}
\end{algorithm}

In our proposed method, the Level of Disorder (\textit{D}) is useful for quantifying and understanding student behavior when answering questions during the assessment.
When $D=0$ , the student answered all questions strictly in a sequence.
However, the higher the value of $D$, the greater the disorder, which provides clues about the learner's difficulties to answer a set of questions. Therefore, this metric must be associated with other metrics such as Student Response Time (\textit{SRT}) or Grouping Deviation (\textit{GD}) to confirm the hypothesis.
Numerical examples and more details are shown in Section~\ref{sec:experiments}.

\subsection{Composition of Other Metrics}

The metrics based on composition of others metrics are: 
Question Comprehension Level, Questionnaire Comprehension Level, and Priority.

\subsubsection{Question Comprehension Level}
The \textbf{Question Comprehension Level (\textit{QCL})} measures the student's learning of the content considering the difficulty level of the question and the time it takes for the student to answer the question.
Thus, \textit{QCL} is based on Question Difficulty Index (\textit{QDI}), Content Difficulty Index (\textit{CDI}) from Table~\ref{tab:DVI}, the Weight of the student's answer (see Section~\ref{sub:WS} from Table~\ref{tab:Weights}) and the Student Response Time (\textit{SRT}). Thereby, the Maximum Comprehension Level (\textit{MCL}) is the highest possible comprehension that can be measured; and the Effective Comprehension Level (\textit{ECL}) effectively measures comprehension, which are obtained by Equation~\ref{eq:MCL} and Equation~\ref{eq:ECL}, respectively.

\begin{equation}\label{eq:MCL}
MCL = QDI\cdot CDI \cdot 4
\end{equation}
\begin{equation}\label{eq:ECL}
ECL = QDI \cdot CDI \cdot w
\end{equation}

The values of $QDI$ and $CDI$ can be obtained from Table~\ref{tab:DVI}, where the value `1' represents the least difficult content or question, `3' is considered a normal difficulty and `5' corresponds to a more difficult concept to understand.

\begin{table}[htbp]
\caption{Difficulty Indices Values}
\begin{center}
\begin{tabular}{|c|c|c|c|}
\hline
\textbf{Index} & \textbf{Easy} & \textbf{Medium} & \textbf{Hard}  \\ \hline
\textbf{Question Difficulty Index - QDI} & 1 & 3 & 5\\ \hline
\textbf{Content Difficulty Index - CDI} & 1 & 3 & 5\\ \hline
\end{tabular}
\end{center}
\label{tab:DVI}
\end{table}

Thereby, $QCL$ considers the time the student took to answer the question, according to the equation proposed in Equation~\ref{eq:QCL}, where we can see that $QCL$ can be calculated in three different ways according to the student response time for each question. In addition, Table~\ref{tab:QCL_1st} demonstrates five case studies.

\begin{equation}
\label{eq:QCL}
QCL = 
\begin{cases}
\frac{ECL}{MCL\cdot 4}, & SRT \leq t/4 \\
\frac{ECL}{MCL}, & t/4 < SRT \leq t\\
\frac{ECL}{MCL + (\frac{SRT - t}{t})}, & SRT > t
\end{cases}
\end{equation}

where $t$ is the \textbf{maximum expected time} for the student to answer the question.

In Table~\ref{tab:QCL_1st}, Q1 implies that the student hit the question, ($w_1=4$), but answered it very quickly ($SRT \leq t/4 \therefore 60 \leq 62.5$), which may indicate that he tried to guess (or cheat) the answer. In this case, by Equation~\ref{eq:QCL}, the $QCL$ will only be 25\%.

On the other hand, Q2 and Q3 were answered within the expected time ($t/4 < SRT \leq t$). That is, ($62.5 < \textbf{70} \leq 250$) and ($45 < \textbf{60} \leq 180$), respectively.
However, only Q2 is completely correct, obtaining a $QCL$ equal to 100\%. Q3 is almost correct, therefore, its \textbf{QCL} is 75\%.
In these specific cases, we can note that the \textbf{QCL} of a student only depends on the weight of his answer.

\begin{table}[htbp]
\scalefont{0.7}
\caption{Question Comprehension Level}
\centering
\begin{tabular}{|c|c|c|c|c|c|c|c|c|c|c|c|}
\hline
\multicolumn{2}{|c|}{\textbf{Question}} & \multicolumn{2}{c|}{\textbf{Q1}} & \multicolumn{2}{c|}{\textbf{Q2}} & \multicolumn{2}{c|}{\textbf{Q3}} & \multicolumn{2}{c|}{\textbf{Q4}} & \multicolumn{2}{c|}{\textbf{Q5}} \\ \hline
\multicolumn{2}{|c|}{\textbf{w}} & \multicolumn{2}{c|}{4} & \multicolumn{2}{c|}{4} & \multicolumn{2}{c|}{3} & \multicolumn{2}{c|}{4} & \multicolumn{2}{c|}{2} \\ \hline
\textbf{CDI} & \textbf{QDI} & 5 & 5 & 3 & 3 & 5 & 5 & 3 & 3 & 5 & 5 \\ \hline
\multicolumn{1}{|l|}{\textbf{ECL}} & \multicolumn{1}{l|}{\textbf{MCL}} & \multicolumn{1}{l|}{100} & \multicolumn{1}{l|}{100} & \multicolumn{1}{l|}{36} & \multicolumn{1}{l|}{36} & \multicolumn{1}{l|}{75} & \multicolumn{1}{l|}{100} & \multicolumn{1}{l|}{36} & \multicolumn{1}{l|}{36} & 50 & 100 \\ \hline
\textbf{SRT} & \textbf{t} & 60 & 250 & 70 & 250 & 60 & 180 & 300 & 250 & 650 & 300 \\ \hline
\multicolumn{2}{|c|}{\textbf{t/4}} & \multicolumn{2}{c|}{62.5} & \multicolumn{2}{c|}{62.5} & \multicolumn{2}{c|}{45} & \multicolumn{2}{c|}{62.5} & \multicolumn{2}{c|}{75} \\ \hline
\multicolumn{2}{|c|}{\textbf{QCL}} & \multicolumn{2}{c|}{25\%} & \multicolumn{2}{c|}{100\%} & \multicolumn{2}{c|}{75\%} & \multicolumn{2}{c|}{99.45\%} & \multicolumn{2}{c|}{49.42\%} \\ \hline
\end{tabular}
\label{tab:QCL_1st}
\end{table}

In the third case, namely, when learners exceed the expected time ($SRT > t \therefore \textbf{300} > 250$), a small value is discounted to distinguish such student from the second case, independent of his answer. Therefore, if this student is more delayed than the prediction, it is likely that their comprehension level is a bit lower than the student that answered in the expected time. Therefore, even though the student answers question Q4 correctly, by Equation~\ref{eq:QCL}, their $QCL$ will be 99.45\%.

Finally, the fifth case shows that the learner also exceeded the expected time ($SRT > t \therefore \textbf{650} > 300$), but answered the question incorrectly. Thus, recieved a weight equal to 2. So, the learner score is calculated according to the third case of Equation~\ref{eq:QCL} and is decreased proportionally to exceeded time. Thus, their $QCL$ will be 49.42\%.

\subsubsection{Questionnaire Comprehension Level}
\label{sec:QuCL}

Unlike the Question Comprehension Level (\textbf{QCL}) metric, the \textbf{Questionnaire Comprehension Level (\textbf{QuCL})} measures the student's learning about the content related to
a questionnaire (or a set of questions). 
This metric is based on \textit{QCL} and the \textit{complement} of the \textbf{Assurance Degree}, that is, $1-AD$ for each question. Equation~\ref{eq:QuCL} details how to calculate the $QuCL$.

\begin{equation}
\label{eq:QuCL}
QuCL = \frac{\sum_{i=1}^{\text{\textbar q\textbar}}QCL_i}{\text{\textbar q\textbar} + (1 - AD)}
\end{equation}

\begin{table}[htbp]
\caption{Questionnaire Comprehension Level}
\centering
\begin{adjustbox}{center, width=\columnwidth-16pt}
\begin{tabular}{|c|c|c|c|c|c|c|}
\hline
\multicolumn{7}{|c|}{\textbf{Subject A}} \\ \hline
\textbf{Question} & \textbf{Q1} & \textbf{Q2} & \textbf{Q3} & \textbf{Q4} & \textbf{Q5} & \textbf{Q6} \\ \hline
\textbf{QCL} & 1 & 0.75 & 1 & 0.50 & 1 & 1 \\ \hline
\textbf{1-AD} & \multicolumn{6}{c|}{0.764} \\ \hline
\multicolumn{7}{|c|}{\textbf{Subject B}} \\ \hline
\textbf{Question} & \textbf{Q1} & \textbf{Q2} & \textbf{Q3} & \textbf{Q4} & \textbf{Q5} & \textbf{Q6} \\ \hline
\textbf{QCL} & 0.75 & 0.75 & 0.75 & 0.75 & 0.75 & 1 \\ \hline
\textbf{1-AD} & \multicolumn{6}{c|}{0.833} \\ \hline
\end{tabular}
\end{adjustbox}
\label{tab:QuCL}
\end{table}

Through Equation~\ref{eq:QuCL} and by using the data from Table~\ref{tab:QuCL}, we can calculate $QuCL$ of the questionnaire, containing the same questions and answers related to Subjects `A' and `B' in Subsection~\ref{sub:WS} (Table~\ref{table:studentGrades}):

$$QuCL_A = \frac{1 + 0.75 + 1 + 0.50 + 1 + 1}{6 + 0.764} \approx 77.61\% $$

$$QuCL_B = \frac{0.75 + 0.75 + 0.75 + 0.75 + 1}{6 + 0.833} \approx 58.54\% $$
In this case, for Subject `A', while $TS_A = 6.67$, and $WS_A = 8.75$, $QuCL_A = 77.61\%$, and for Subject `B', while $TS_B=1.67$ and $WS_B=7.92$, we have a $QuCL_B=58.54\%$, where we not only consider the weights, hits and errors, but also the difficulty level of each question.
We see that this metric uses elements of almost all the metrics defined so far and can more accuratley demonstrate student's understanding than Traditional Score (TS).

\subsubsection{Priority}

\textbf{Priority (\textbf{P})} (see Equation~\ref{eq:priority})  indicates the relevance of a subject (or topic) to be studied by a student based on the scores obtained by the Traditional Score (\textbf{TS}) and Weighted Score (\textbf{WS}).

\begin{equation}
\label{eq:priority}
P = (10 - TS)\cdot \frac{WS}{10}
\end{equation}

As previously shown (Subsection~\ref{sub:WS}), for Subject ``A'' we have $TS = 6.67$ and $WS = 8.75$; and for Subject ``B'', $TS = 1.67$ and $WS = 7.92$. Therefore, we can use Equation~\ref{eq:priority} to calculate the priority for each subject resulting in:

$$P_A = (10 - 6.67)\cdot \frac{8.75}{10} = 2.914$$
$$P_B = (10 - 1.67)\cdot \frac{7.92}{10} = 6.597$$

Where, the priority for Subject ``A'' is $2.914$, and $6.597$ for Subject ``B''. Observing 
Table~\ref{table:studentGrades}, we can see that despite the better overall performance of Subject A, Subject B has more priority and, therefore, is recommended to study more. This happens because the student \textit{almost} hit more questions in Subject B, which means that a low  level of effort for learning and reaching a high performance is needed.

\section{Experiments}
\label{sec:materials}

To demonstrate our proposed metrics, we developed a simple educational platform that includes a multiple-choice web-based questionnaire containing five possible options for each question. 
Therefore, students can select or change their answers as many times as they want before giving their final answer.
There are also navigation buttons that allow students to review their answers.
As presented in Section~\ref{sec:introduction}, this platform collects some additional data at runtime, such as the time students spend on each question and how many times the student changed an answer to the same question.
Details about this educational platform are beyond the scope of this paper.

The data used in this experiment is restricted to one 10th grade high school class with 33 students.
We implemented a questionnaire with 40 questions including the following subjects: Portuguese, English and Spanish Languages, History, Chemistry, Physics, Sociology, Mathematics, Geography and Biology.
Four hours was the maximum time given to answer all questions.

The interaction of the students with the questionnaire was recorded in a {\it log} file in CSV (Comma Separated Values) format, where the question ID and timestamp of each event were recorded. Thus, the set of events includes the selected questions and all chosen answers. The metrics were calculated in Python and stored in a NoSQL database using MongoDB.
\section{Results and Discussions}
\label{sec:experiments}

This section presents some extracts from collected data, aiming to demonstrate the possibilities offered by the proposed metrics. 

\subsection{Metrics Analysis}
\label{subsection:class_analysis}

\subsubsection{Assessing Students}
In order to analyze the metrics, we used a frequency distribution method to classify the students into a set of groups (see Table~\ref{tab:Groups}), where the \textbf{Floor} value is a closed interval and the \textbf{Ceiling} value is an open interval.
The total number of groups is given by the square root of total students, that is $k = \sqrt{33} \approx 6$. Considering that the groups should be in the interval $[0, 1]$, the amplitude $h$ for each group is given by $h = (1/k) = (1/6) = 0.1667$.



\begin{table}[htbp]
\caption{Groups of Students}
\centering
\begin{tabular}{|c|c|c|}
\hline
\multicolumn{1}{|l|}{\textbf{Groups}} & \multicolumn{1}{l|}{\textbf{Floor}} & \multicolumn{1}{l|}{\textbf{Ceiling}} \\ \hline
\textbf{Group 1} & 0.0000 & 0.1667 \\ \hline
\textbf{Group 2} & 0.1667 & 0.3333 \\ \hline
\textbf{Group 3} & 0.3333 & 0.5000 \\ \hline
\textbf{Group 4} & 0.5000 & 0.6667 \\ \hline
\textbf{Group 5} & 0.6667 & 0.8333 \\ \hline
\textbf{Group 6} & 0.8333 & 1.0000 \\ \hline
\end{tabular}
\label{tab:Groups}
\end{table}



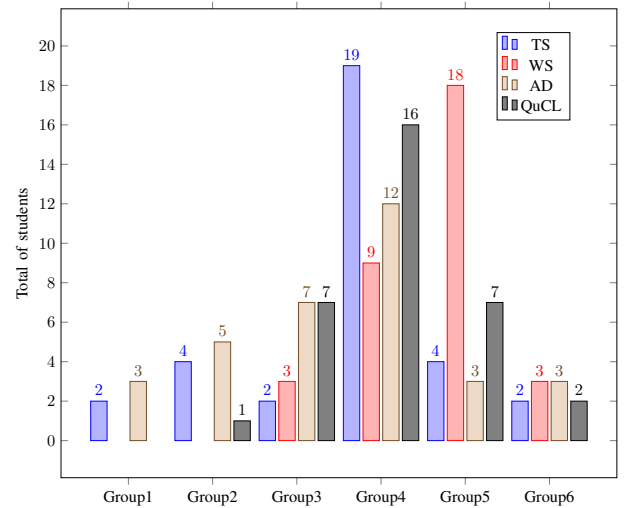
\begin{figure}[htb]
	\centering
	\begin{tikzpicture}[scale=0.62]
\pgfplotsset{width=13.5cm,compat=1.9}
\begin{axis}[
    ybar,
    enlargelimits=0.16,
    legend style={at={(0.85,0.95)},
      anchor=north,legend columns=1},
    ylabel={Total of students
},
    symbolic x coords={Group1, Group2, Group3, Group4, Group5, Group6},
    xtick=data,
    nodes near coords,
    nodes near coords align={vertical},
    ]

\addplot+[y filter/.expression={y==0 ? nan : y}] coordinates {
(Group1,2) 
(Group2,4) 
(Group3,2) 
(Group4,19) 
(Group5, 4) 
(Group6, 2)};

\addplot+[y filter/.expression={y==0 ? nan : y}] coordinates {
(Group1,0)
(Group2,0) 
(Group3,3) 
(Group4,9) 
(Group5, 18)
(Group6, 3)};

\addplot+[y filter/.expression={y==0 ? nan : y}] coordinates {
(Group1,3)
(Group2,5) 
(Group3,7)
(Group4,12)
(Group5,3)
(Group6,3)};

\addplot+[y filter/.expression={y==0 ? nan : y}] coordinates {
(Group1,0)
(Group2,1)
(Group3,7)
(Group4,16) 
(Group5,7)
(Group6,2)};
\legend{TS, WS, AD, QuCL}
\end{axis}
\end{tikzpicture}
\caption{Students organized by groups of scores}
     \label{fig:groups_students}
\end{figure}

Figure~\ref{fig:groups_students} shows the total number of students in each group according to \textit{TS}, \textit{WS}, \textit{AD} and \textit{QuCL} metrics, which describe the students' performance on a set of questions.
If we add up all the bars of all metrics, the value is always 33.
When considering \textit{TS}, \textit{AD} and \textit{QuCL} metrics, most students are in the groups 3-4, and consequently, their scores are between 0.3333 and 0.6667
(see Groups 3 and 4 in Table~\ref{tab:Groups}).
On the other hand, when considering the WS metric, most students are in the groups 4-5,
and consequently, their scores are between 0.5000 and 0.8333 (see Groups 4 and 5 in Table~\ref{tab:Groups}), indicating that this metric gave a higher score than the others.
However, it is important to note the \textit{TS} metric includes most students (19 people) in Group 4 (from 0.5000 to 0.6667); and \textit{QuCL} and \textit{AD} show a curve similar to a normal distribution.

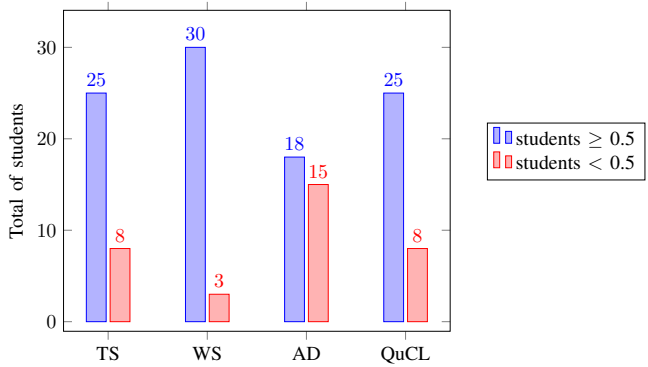
\begin{figure}[ht]
	\centering
	\begin{tikzpicture}[scale=0.75]
\begin{axis}[
    ybar,
    enlargelimits=0.15,
    legend style={at={(1.3,0.65)},
      anchor=north,legend columns=1},
    ylabel={Total of students
},
    symbolic x coords={TS,WS,AD, QuCL},
    xtick=data,
    nodes near coords,
    nodes near coords align={vertical},
    ]
\addplot coordinates {(TS,25) (WS,30) (AD,18) (QuCL,25)};
\addplot coordinates {(TS,8) (WS,3) (AD,15) (QuCL,8)};
\legend{students $\geq$ 0.5, students $<$ 0.5}
\end{axis}
\end{tikzpicture}
	\caption{Grouped students by score situation}
     \label{fig:students_media}
\end{figure}

Figure~\ref{fig:students_media} presents the quantity of students who recieved the minimum score for approval in Brazilian high school, i.e., a score equal to or higher than 0.5. Firstly, we confirm the tendency of \textit{WS} to give a higher score than other metrics, since thirty students obtained a score higher than 0.5. This occurs because this metric is based on the weight of the answer, which gives some points when the answer is close to the correct one.
Almost half of the students were not completely sure of their answers (\textit{AD} metric). However, most of the students have high comprehension levels (\textit{QuCL} metric).


\subsubsection{Assessing Subjects}


\begin{table}[htbp]
\caption{Priorities of Subjects}
\centering
\begin{tabular}{|l|c|}
\hline
\textbf{Subject} & \textbf{Priority} \\ \hline
\textbf{English Language} & 0.393 \\ \hline
\textbf{Geography} & 0.373 \\ \hline
\textbf{Portuguese Language} & 0.322 \\ \hline
\textbf{Biology} & 0.288 \\ \hline
\textbf{Physics} & 0.260 \\ \hline
\textbf{Chemistry} & 0.257 \\ \hline
\textbf{Maths} & 0.240 \\ \hline
\textbf{History} & 0.205 \\ \hline
\end{tabular}
\label{tab:PrioritiesSubject}
\end{table}

\begin{table}[htbp]
\caption{Priorities of Subject: Geography}
\centering
\begin{tabular}{|l|l|}
\hline
\textbf{Earth Dynamics} & 0.402 \\ \hline
\textbf{Scale} & 0.384 \\ \hline
\textbf{Cartography} & 0.325 \\ \hline
\end{tabular}
\label{tab:PrioritiesBiology}
\end{table}

Furthermore, after calculating and ordering the average of the priorities from all students in the class, Table~\ref{tab:PrioritiesSubject}
(normalized in the range $[0,1]$)
shows that three subjects with the highest priority are English Language (39.3\%), Geography (37.3\%) and Portuguese Language (32.2\%), respectively. Thus, it can calculate the \textbf{priority topics} for each teacher to review in the classroom with all students. For instance, in relation to the subject Geography, Table~\ref{tab:PrioritiesBiology} presents ``Earth Dynamics'' as the highest priority topic, followed by ``Scale'' and ``Cartography'', respectively.

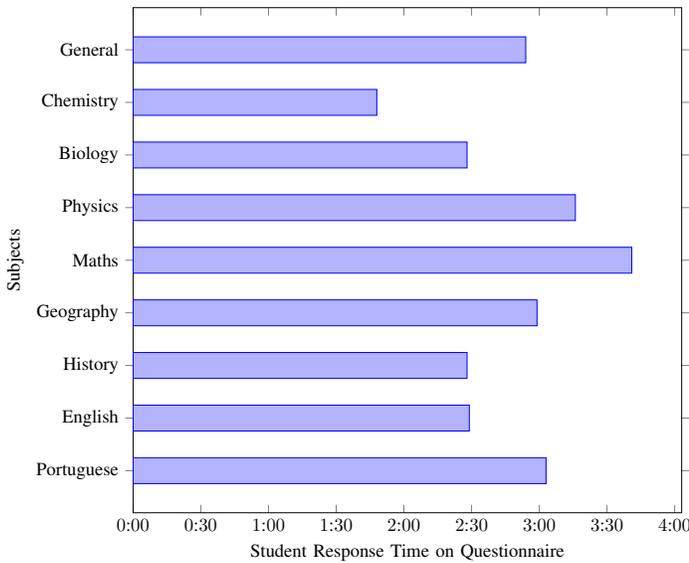
\begin{figure}[htb]
		\centering
		\begin{tikzpicture}[scale=0.7]
		\begin{axis}[height=10cm, width=12cm,
		xbar,
		ylabel=Subjects,
		xlabel=Student Response Time 
		on Questionnaire,
		symbolic y coords={Portuguese,
			English,
			History,
			Geography,
			Maths,
			Physics,
			Biology,
			Chemistry,
			General},
		bar width=14pt,
		y=1.0cm,
		x filter/.code={\pgfmathparse{#1/60}},
		xticklabel={ 
			\pgfmathsetmacro\hours{floor(\tick)}%
			\pgfmathsetmacro\minutes{(\tick-\hours)*0.6}%
			\pgfmathprintnumber{\hours}:\pgfmathprintnumber[fixed, fixed zerofill, skip 0.=true, dec sep={}]{\minutes}%
		},
		xmin=0,
		]
		\addplot table{
		num values
		183 Portuguese
		149 English
		148 History
		179 Geography
		221 Maths
		196 Physics
		148 Biology
		108 Chemistry
		174 General
		};
		\end{axis}
		\end{tikzpicture}
		\caption{Average time spent in questions of each subject}
		\label{fig:students_time_questionnaire}
	\end{figure}
	
Figure~\ref{fig:students_time_questionnaire} shows the average time spent on each subject, according to \textit{SRT} Metric (Section~\ref{sec:SRT}). We can see that Math, Physics and Portuguese Language were the subjects where students spent the most time. All these subjects are above the general class average.
In relation to Portuguese Language, Table~\ref{tab:SRTPortuguese} compares the expected answer time to the average effective answer time of all students, where only questions 4, 6 and 7 presented effective times that were lower than the expected time.
This situation slightly decreases the \textit{QuCL} metric since students fit the third case of Equation~\ref{eq:QCL}, i.e, $SRT > t$.

\begin{table}[htbp]
\caption{Time spent in Portuguese Language}
\centering
\begin{tabular}{|c|c|c|}
\hline
\textbf{Questions} & \textbf{SRT Average} & \textbf{Expected Time (t)} \\ \hline
\textbf{Q1} & 04:47 & 03:20 \\ \hline
\textbf{Q2} & 02:50 & 02:00 \\ \hline
\textbf{Q3} & 02:58 & 01:20 \\ \hline
\textbf{Q4} & \textbf{02:17} & \textbf{03:00} \\ \hline
\textbf{Q5} & 02:49 & 01:20 \\ \hline
\textbf{Q6} & \textbf{02:18} & \textbf{03:00} \\ \hline
\textbf{Q7} & \textbf{02:46} & \textbf{03:00} \\ \hline
\textbf{Q8} & 03:38 & 03:00 \\ \hline
\textbf{Total} & 24:22 & 20:00 \\ \hline
\end{tabular}
\label{tab:SRTPortuguese}
\end{table}

As presented in Section~\ref{sec:level-disorder} (Level of Disorder), when disorder is equal to zero ($D=0$) this means that the student answered in a rigid way, from the first to the last question.
Table~\ref{tab:LevelDisorderClass} presents the Level of Disorder ($D$) metric considering all subjects. The Level of Disorder is a number in the range $[0,1]$, but in that table it is shown as a percentage. The columns are: (i) the average disorder of each subject including all students; and (ii) the percentage of students with non-zero disorder in relation to class size.

Thus, the column ``\textbf{Average (all)}" shows that the highest disorder considering all students was the subject Mathematics, with an average of 32.52\%. Besides that, in column ``\textbf{Percent.($D>0$)}", Mathematics has the highest proportion of students with a level of disorder, i.e., 69.70\%.
This result could indicate that students may have more difficulties with Maths. This can be corroborated by Figure~\ref{fig:students_time_questionnaire}, which shows that students spend more time on Maths than on other subjects.
However, even though almost 70\% of the students answered the questions out of order in Maths, its level of disorder can be considered low since the average disorder level considering all students is $32.52\%$.
This information should be considered in the analysis.


\begin{table}[htbp]
\caption{Levels of Disorder of Subjects}
\centering
\begin{tabular}{|l|c|c|}
\hline
\textbf{Subjects} & \textbf{Average (all)} &  \textbf{Percent.($D>0$)} \\ \hline
Portuguese L. & 25.04\% & 54,55\% (18) \\ \hline
English L. & 13.44\% & 24,24\% (8)\\ \hline
History & 31.55\% & 54,55\% (18)\\ \hline
Geography & 17.45\% & 33,33\% (11)\\ \hline
Mathematics & 32.52\% & 69,70\% (23)\\ \hline
Physics & 18.54\% & 36,36\% (12)\\ \hline
Biology & 29.62\% & 51,52\% (17)\\ \hline
Chemistry & 31.40\% & 54,55\% (18) \\ \hline
\end{tabular}
\label{tab:LevelDisorderClass}
\end{table}

In addition, we could compare the level of disorder with other metrics such as Assurance Degree (\textit{AD}) and 
Student Response Time (\textit{SRT}) to verify if the high disorder is due to student confusion (low assurance degree) or because they did not know the answers (high grouping deviation or time spent). 


For example, regarding Mathematics,  Figure~\ref{fig:students_time_questionnaire} shows that the average time spent to answer a question is higher than all others subjects. We can also see in Tab.~\ref{tab:ADGDClass} that the average Assurance Degree ($AD$) of the class is only 36.29\%.
This implies that students had more difficulty in this subject than in others, as they spent more time and were less sure of their answers with high deviation in relation to correct answers.

\begin{table}[htbp]
\caption{Assurance Degree of Subjects}
\centering
\begin{tabular}{|l|c|}
\hline
\multicolumn{1}{|c|}{\textbf{Subject}} & \textbf{AD} \\ \hline 
Portuguese Language & 0.4825 \\ \hline 
English Language & 0.3755 \\ \hline 
History & 0.2446 \\ \hline 
Geography & 0.4162 \\ \hline 
Maths & 0.3629 \\ \hline 
Physics & 0.5061 \\ \hline 
Biology & 0.4466 \\ \hline 
Chemistry & 0.2611 \\ \hline 
General & 0.3831 \\ \hline 
\end{tabular}
\label{tab:ADGDClass}
\end{table}

\subsection{Metrics Composition: Assurance Degree vs Questionnaire Comprehension Level}
\label{subsection:AD_QuCL}

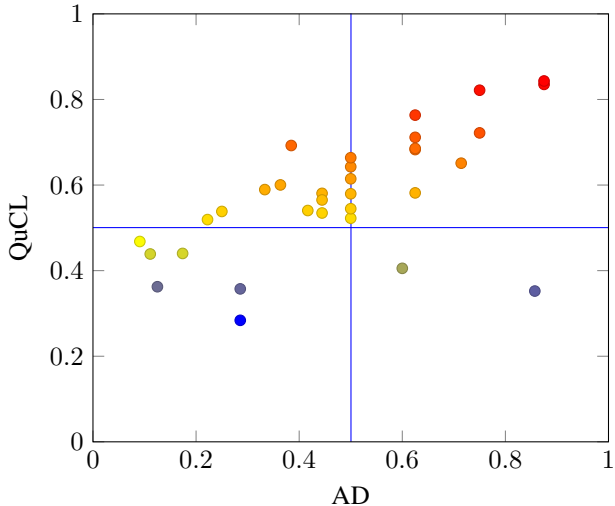
\begin{figure}[htb]
\centering
\begin{tikzpicture}
\draw[blue] (0,2.85) -- (6.85,2.85);
\draw[blue] (3.43,0) -- (3.43,5.7);
    \begin{axis}[
    xmin=0, xmax=1,
    ymin=0, ymax=1,
    xlabel={AD},
    ylabel={QuCL}]   
    \addplot[scatter, only marks] 
    table[x=x,y=y]{
        x    y
    0.2857	0.2839
    0.3846	0.6924
    0.5000	0.6146
    0.7500	0.8216
    0.6250	0.6829
    0.0909	0.4681
    0.8750	0.8352
    0.4444	0.5806
    0.8750	0.8432
    0.3333	0.5894
    0.5000	0.5796
    0.6000	0.4054
    0.4444	0.5347
    0.6250	0.7115
    0.6250	0.5817
    0.3636	0.6003
    0.5000	0.5225
    0.7143	0.6511
    0.4167	0.5405
    0.7500	0.7219
    0.6250	0.6855
    0.2222	0.5193
    0.2500	0.5383
    0.1739	0.4402
    0.1111	0.4389
    0.6250	0.7634
    0.1250	0.3624
    0.4444	0.5652
    0.5000	0.5447
    0.8571	0.3523
    0.5000	0.6427
    0.2857	0.3574
    0.5000	0.6639
    };
    \end{axis}
\end{tikzpicture}
     \caption{AD vs. QuCL Comparison - Portuguese Language}
     \label{fig:TAS_QuCL}
 \end{figure}

In order to determine the metrics Assurance Degree ($AD$) and Questionnaire Comprehension Level ($QuCL$), we clustered the results of both metrics into four groups. The chart in Figure~\ref{fig:TAS_QuCL} shows the composition. Therein, we plotted two perpendicular lines at the points 0.5 on each axis, as this is the grade established for student approval in most Brazilian high-schools.

Considering the concept of Cartesian Plane quadrants, the first quadrant contains students above or equal to 0.5 in both metrics, i.e., these students have good performance both in $QuCL$ and in $AD$ and, because of this, we do not need to worry when compared to students in the others quadrants. For example, the three students with $QuCL$ greater than 0.8, have an $AD$ greater than 0.75. 
This means that they have very high comprehension and confidence. 


On the other hand, in the third quadrant, we note students in an opposite situation, i.e. with $QuCL$ and $AD$ below 0.5. This situation is more worrying because students have both low comprehension and assurance levels. In Figure~\ref{fig:TAS_QuCL}, all six students in this quadrant have an AD below 0.3, i.e. they have an Assurance Degree below 30\%. A more critical case of $QuCL$ is one student with 0.286. When analysing both metrics, the students with more difficulties are the three students with $AD$ between 0.1 and 0.3 and $QuCL$ between 0.2 and 0.4.
 
The other quadrants have students with different behaviors and needs. For example, in the second quadrant, there are students with high $QuCL$ (above 50\%) and low confidence ($AD$ below 50\%).
This means that these students are getting the questions right, but still have a lot of doubts.
However, even though these students have many doubts these results show that they are able to self-correct.
 
Finally, the fourth quadrant presents students with high degrees of confidence, but low levels of comprehension. This means that students could have answered the questions incorrectly or very quickly, possibly trying to guess. To better understand each behavior, it is necessary to verify the metrics of each specific question or set of questions.

Thus, in the next section, we present a detailed analysis of the most critical case. 
 
\subsection{Analyzing students from the Third Quadrant}\label{subsection:3rdQnd}

As stated in Section~\ref{subsection:AD_QuCL}, Figure~\ref{fig:TAS_QuCL} shows that students in the third quadrant have low comprehension and assurance levels and need more support. Therefore, we focus on some of them to demonstrate how the information from new metrics can be used to help overcome their difficulties.

\begin{figure}[htb]
		\centering
		\begin{tikzpicture}[scale=0.54]
		\pgfplotsset{width=17cm,compat=newest}
		\begin{axis}[
		ybar,
		xtick=data,
		legend style={at={(0.92,0.95)},
			anchor=north,legend columns=1},
		symbolic x coords={0322, 0290, 0304, 0320, 0398, ClassAvg.},
		enlarge x limits={abs=1.6cm},
		bar width = 2.00mm,
		ylabel=Time (MM:SS),
		xlabel=Students,
		x=2.3cm,
		y filter/.code={\pgfmathparse{#1/60}},
		yticklabel={ 
			\pgfmathsetmacro\hours{floor(\tick)}%
			\pgfmathsetmacro\minutes{(\tick-\hours)*0.6}%
			\pgfmathprintnumber{\hours}:\pgfmathprintnumber[fixed, fixed zerofill, skip 0.=true, dec sep={}]{\minutes}%
		},
		xticklabel style=
		{rotate=0,anchor=near xticklabel},
		ytick={0,1.05,...,10.55},
		]
		\addplot table {
			num value
			0322	38
			0290	148
			0304	206
			0320	258
			0398	255
			ClassAvg.	286
			
		};
		\addplot table {
			num value
			0322		322
			0290		16
			0304		310
			0320		63
			0398		333
			ClassAvg.		170
			
		};
		\addplot table {
			num value
			0322			0
			0290			95
			0304			284
			0320			90
			0398			0
			ClassAvg.			178

		};
		\addplot table {
			num value
			0322				345
			0290				37
			0304				169
			0320				77
			0398				141
			ClassAvg.				136		
		};
		\addplot table {
			num value
			0322				586
			0290				227
			0304				1
			0320				108
			0398				99
			ClassAvg.				168
			
		};
		\addplot table {
			num value
			0322					7
			0290					571
			0304					111
			0320					80
			0398					159
			ClassAvg.					137
			
		};
		\addplot[black!20!black, fill=red] table {
			num value
			0322						54
			0290						60
			0304						259
			0320						91
			0398						148
			ClassAvg.						165						
		};
		\addplot[fill=yellow] table {
			num value
			0322							298
			0290							113
			0304							51
			0320							375
			0398							350
			ClassAvg.							217	
		};
		\legend{1, 2, 3, 4, 5, 6, 7, 8}	
		\end{axis}
		\end{tikzpicture}
		\caption{Time spent by students with lowest AD}
		\label{fig:students_time}
	\end{figure}
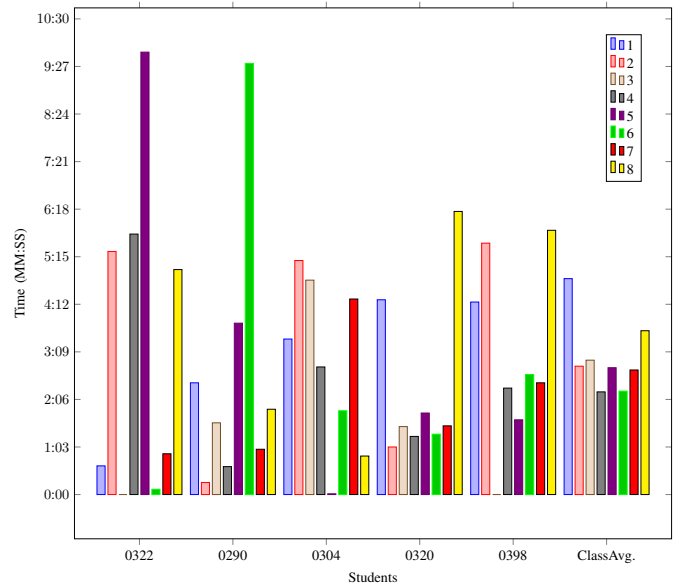

In Table~\ref{tab:3rdQstudents}, we compared the metrics of the five students with the lowest $QuCL$ of 3rd quadrant, i.e., values between 0.284 and 0.440. 
All data in Table~\ref{tab:3rdQstudents} were normalized between 0 and 1 to make it easy to compare metrics. 
Furthermore, Figure~\ref{fig:students_time} shows the time spent by students on each question of Portuguese Language, where the evaluated topics are available in Table~\ref{tab:3rdQstudents_Topic}.

\begin{table}[htbp]
\caption{students with lowest QucL}
\centering
\begin{tabular}{|c|c|c|c|c|c|}
\hline
\textbf{Student ID} & \textbf{TS} & \textbf{WS} & 
\textbf{AD} & \textbf{QuCL} & \textbf{Disorder} \\ \hline
0322 & 0.250 & 0.344 & 
0.286 & 0.284 & 0.000 \\ \hline
0290 & 0.500 & 0.688 & 
0.174 & 0.440 & 0.410 \\ \hline
0304 & 0.375 & 0.531 & 
0.111 & 0.439 & 0.349 \\ \hline
0320 & 0.125 & 0.406 & 
0.125 & 0.362 & 0.000 \\ \hline
0398 & 0.250 & 0.406 & 
0.286 & 0.357 & 0.000 \\ \hline
\end{tabular}
\label{tab:3rdQstudents}
\end{table}

Table~\ref{tab:3rdQstudents} shows that for this group of students, TS is between 0.125 and 0.500, $WS$ is between 0.344 and 0.688, 
AD is between 0.111 and 0.286 and just two students presented Disorder above 0\%. 
It is worth noting that a low assurance degree may have occurred not because students frequently changed their answers, but because they missed many questions.

We can confirm this hypothesis by simply looking at the total number of students that got it right. In this case, the data from Table~\ref{tab:3rdHitsMisses} shows that no one hit questions 5 and 7, only one student got the questions 1, 3, 6, and 8 correct, but four students got questions 2 and 4 correct. However, only using $TS$ metric (common hits) we cannot confirm which topics students must prioritize. In view of this, we calculated the $P$ metric for this questionnaire in Tables~\ref{tab:3rdQstudents_Priority} and~\ref{tab:Portuguese_PriorityClass}.

\begin{table}[htbp]
\caption{Hits (1) and misses (0) - Portuguese Language}
\centering
\begin{tabular}{|c|c|c|c|c|c|c|c|c|}
\hline
\multicolumn{1}{|c|}{\textbf{Students}} & \textbf{Q1} & \textbf{Q2} & \textbf{Q3} & \textbf{Q4} & \textbf{Q5} & \textbf{Q6} & \textbf{Q7} & \textbf{Q8} \\ \hline
0322 & 0 & 0 & 0 & 1 & 0 & 0 & 0 & 1 \\ \hline
0290 & 1 & 1 & 1 & 1 & 0 & 0 & 0 & 0 \\ \hline
0304 & 0 & 1 & 0 & 1 & 0 & 1 & 0 & 0 \\ \hline
0320 & 0 & 1 & 0 & 0 & 0 & 0 & 0 & 0 \\ \hline
0398 & 0 & 1 & 0 & 1 & 0 & 0 & 0 & 0 \\ \hline
Total Hits & 1 & 4 & 1 & 4 & 0 & 1 & 0 & 1 \\ \hline
\end{tabular}
\label{tab:3rdHitsMisses}
\end{table}

\begin{table}[htbp]
\caption{Topics of Portuguese Language}
\centering
\begin{tabular}{|c|l|c|}
\hline
\textbf{ID Topic} & \multicolumn{1}{c|}{\textbf{Description}} & \textbf{Question} \\ \hline
17 & Text interpretation & 4, 7, 8 \\ \hline
81 & Orthography & 3\\ \hline
91 & Linguistic Variation & 1, 2 \\ \hline
92 & Verb & 5, 6\\ \hline
\end{tabular}
\label{tab:3rdQstudents_Topic}
\end{table}

Table~\ref{tab:3rdQstudents_Topic} shows the topic ID, its description, and which question evaluated this topic. In Table~\ref{tab:3rdQstudents_Priority}, the priority of the topic calculated for each student is presented in descending order. So, students  ``0322'', ``0290'', and ``0320'' must prioritize studying ``Text Interpretation''; student ``0304" has to focus on ``Linguistic Variation'; and student ``0398" has to focus on ``Verb''. Afterwards, the second priority study area is suggested to students, and so on.

\begin{table}[htbp]
\caption{Priority of Topics.}
\centering
\begin{tabular}{|c|l|c|l|c|l|c|l|c|}
\hline
\textbf{ID student} & \multicolumn{2}{c|}{\textbf{Priority1}} & \multicolumn{2}{c|}{\textbf{Priority2}} & \multicolumn{2}{c|}{\textbf{Priority3}} & \multicolumn{2}{c|}{\textbf{Priority4}} \\ \hline
0322 & 17 & 0.278 & 91 & 0.125 & 81 & 0.000 & 81 & 0.000 \\ \hline
0290 & 17 & 0.389 & 92 & 0.375 & 91 & 0.000 & 91 & 0.000 \\ \hline
0304 & 91 & 0.438 & 92 & 0.313 & 81 & 0.250 & 17 & 0.222 \\ \hline
0320 & 17 & 0.500 & 91 & 0.313 & 81 & 0.250 & 92 & 0.125 \\ \hline
0398 & 92 & 0.375 & 17 & 0.333 & 91 & 0.250 & 81 & 0.000 \\ \hline
\end{tabular}
\label{tab:3rdQstudents_Priority}
\end{table}

Aiming to best support the teachers in terms of reinforcement classes for all students, Table~\ref{tab:Portuguese_PriorityClass} shows the priority of each topic for the class (not for a specific student). In this case, the topic with highest priority is  ``Linguistic Variation'', followed by ``Verb'', ``Text Interpretation'' and, finally ``Orthography''.

\begin{table}[htbp]
\caption{Class - Priority of Topics}
\centering
\begin{tabular}{|l|l|l|l|l|l|l|l|}
\hline
\multicolumn{2}{|c|}{\textbf{Priority1}} & \multicolumn{2}{c|}{\textbf{Priority2}} & \multicolumn{2}{c|}{\textbf{Priority3}} & \multicolumn{2}{c|}{\textbf{Priority4}} \\ \hline
91 & 0.342 & 92 & 0.329 & 17 & 0.323 & 81 & 0.263 \\ \hline
\end{tabular}
\label{tab:Portuguese_PriorityClass}
\end{table}

Considering the time spent by these students and analyzing Figure~\ref{fig:students_time}, we can see that some students answered questions faster than the class average. In some cases, we might consider this a guess. See student 0322- questions 3 and 6; student 0290- question 2; student 0304- question 5; and student 0398- question 3. 

By comparing the Table~\ref{tab:SRTPortuguese} and Figure~\ref{fig:students_time}, we can see that  student ``0322'' exceeded the time for Question 2 by 03:22 min and the student ``0398'' by 03:33 min. Similarly, they exceeded the time for Question 8 by 01:58 and 02:50 min, respectively.

From the data of these students for these specific questions, student ``0322'' hit Q8 but missed question 2, and the opposite happened with student ``0398''. Thus, the extra time spent on these questions was only helpful for one of them. Besides that, from Table~\ref{tab:QD_3rdquadrant}, the Question Doubt ($QD$) for both questions and students was equal to zero, indicating that they thought about the answer for the whole time and only answered once. Despite the low value of $QD$ for these questions, the Assurance Degree (measured for the whole questionnaire) is low for both students because, in general, they missed most of questions according to $TS$ metric.

\begin{table}[htbp]
\caption{Question Doubt - Students with lowest QuCL}
\centering
\begin{tabular}{|c|c|c|c|c|c|c|c|c|}
\hline
\textbf{Student} & \textbf{Q1} & \textbf{Q2} & \textbf{Q3} & \textbf{Q4} & \textbf{Q5} & \textbf{Q6} & \textbf{Q7} & \textbf{Q8} \\ \hline
0322 & 0 & 0 & -1 & 0 & 0 & 0 & 0 & 0 \\ \hline
0290 & 1 & 0 & 0 & 0 & 3 & 10 & 1 & 0 \\ \hline
0304 & 13 & 0 & 1 & 0 & 0 & 1 & 3 & 1 \\ \hline
0320 & 0 & 0 & 0 & 0 & 0 & 0 & 0 & 0 \\ \hline
0398 & 0 & 0 & -1 & 0 & 0 & 0 & 0 & 0 \\ \hline
\end{tabular}
\label{tab:QD_3rdquadrant}
\end{table}
\section{Conclusion}\label{sec:conclusions}

Assessment is the systematic process of using empirical data to measure knowledge, skills, attitudes, and beliefs. During assessment, teachers try to improve a student's path toward learning. On the other hand, the evaluation focuses on grades and may reflect various situations within the classroom, as well as what was actually learned. Therefore, assessment is adopted to measure student learning, while evaluation is more often applied to measure performance. In this case, any metric can be adopted to measure either learning or performance.

In this paper, we propose several metrics for student assessment. The main purpose of these metrics is not only to assess student learning of the studied content, but also to provide opportunities for teachers to improve student learning. 
However, although we present some learning analytics based on the proposed metrics, the analysis, inferences, and recommendations are outside the scope of this paper.

Most educational institutions still adopt the traditional score, namely, based on hits or mistakes as a parameter to measure student learning. However, we argue that traditional score is not enough to correctly assess student’s performance. Thus, this paper presents some new metrics to assess learning and give feedback to students.
Despite the several kinds of assessment, we focus on formative and summative assessments, which primarily aim to educate and improve student learning.

The proposed metrics are divided in two parts: (i) isolated metrics (Weighted Score, Question Doubt, Assurance Degree, Student Response Time, and Level of Disorder); and (ii) metrics based on composition of other metrics (Question Comprehension Level, Questionnaire Comprehension Level, and Priority).

The Weighted Score ($WS$) is a learning metric that is different from Traditional Score ($TS$) as it considers when a student's answer is wrong but close to the correct answer. Thus, to determine scores based only on right/wrong answers is not enough to measure student's understanding of a subject nor to help improve the student's path toward learning. The Weighted Score could help dismiss such discrepancies. 
%
%
Question Doubt ($QD$) is a metric that calculates the number of times a student returns to the same question and changes it. This metric is important for determining the doubt level of a student in a question, which could show that a student is not confident in a topic or concept of a given question.
Assurance Degree ($AD$) measures the self-confidence a student has when answering a set of questions. This metric relies on the total number of correct answers and the number of times the student answered the questions. Using this metric, we can estimate the level of assurance of a student, and if a student changes their answers often, their assurance level will be low. This metric is usually used in combination with other metrics.
Student Response Time ($SRT$) is a metric that measures the accumulated time spent to answer each question. This metric is very useful as it can provide information about student's difficulties to answer a question, when the student may not be confident with the topic/concept or if the question has a high level of difficulty. 
The Level of Disorder ($D$) is a metric that uses the disorder degree of the events. We assume that teachers present questions in order, from first to last. The Level of Disorder ($D$) can quantify and understand student behavior while answering questions during the assessment.

The Question Comprehension Level and Questionnaire Comprehension Level are composed of other metrics. The Question Comprehension Level ($QCL$) measures student’s learning about a content considering the difficulty level of the question, and the Student Response Time ($SRT$), which is the time it takes the student to answer that question. In this case, we consider three cases: (i) when a student answered faster; (ii) when a student answered in the time determined by the teacher; and (iii) when a student answered after the expected time. On the other hand, the Questionnaire Comprehension Level ($QuCL$) measures student learning of a content related to a questionnaire (or a set of questions). This metric is based on Question Comprehension Level and the Assurance Degree.

In order to give feedback to students, we propose the Priority ($P$) metric, which prioritizes a subject or topic to be studied by the students. 

We present the results of an experiment including 33 10th grade high school students. Therefore, we describe all calculated metrics of the data collected from a 40-question questionnaire. When presenting the metrics, it was possible to see feedback possibilities for students and teachers. 

Another outcome was a clustering-based method based on the composition of two specific metrics, namely, Assurance Degree ($AD$) and Questionnaire Comprehension Level ($QuCL$). In this case, we clustered both metrics into four groups considering the grade established for student approval( 50\%) in most Brazilian high-schools. The results were plotted in four quadrants of a Cartesian Plane. The first quadrant contains students above or equal to 50\% in both metrics, i.e., students with good performance both in $QuCL$ and $AD$. On the other hand, in the third quadrant, we note students in an opposite situation, i.e. with $QuCL$ and $AD$ below 50\%. This is a more worrisome situation because these students present both low comprehension and assurance levels. The specific students in the third quadrant were analysed to discover why they presented poor performance. We conclude that such data is useful for feedback and recommendations.

Regarding future research, we have a lot of data and there are many machine learning methods that could provide several learning analytics.
Another future goal is to automatically provide these methods, based on the metrics, feedback and recommendations to both students and teachers.

\section*{Acknowledgements}
This research, according to Article 48 of Decree nº 6.008/2006, was funded by Samsung Electronics of Amazonia Ltda, under the terms of Federal Law nº 8.387/1991, through agreement nº 003, signed by  ICOMP/UFAM, and by Amazonas State Research Support Foundation (FAPEAM) through project 122/2018 (Universal) and PAPAC. 

\bibliography{references}

\end{document}